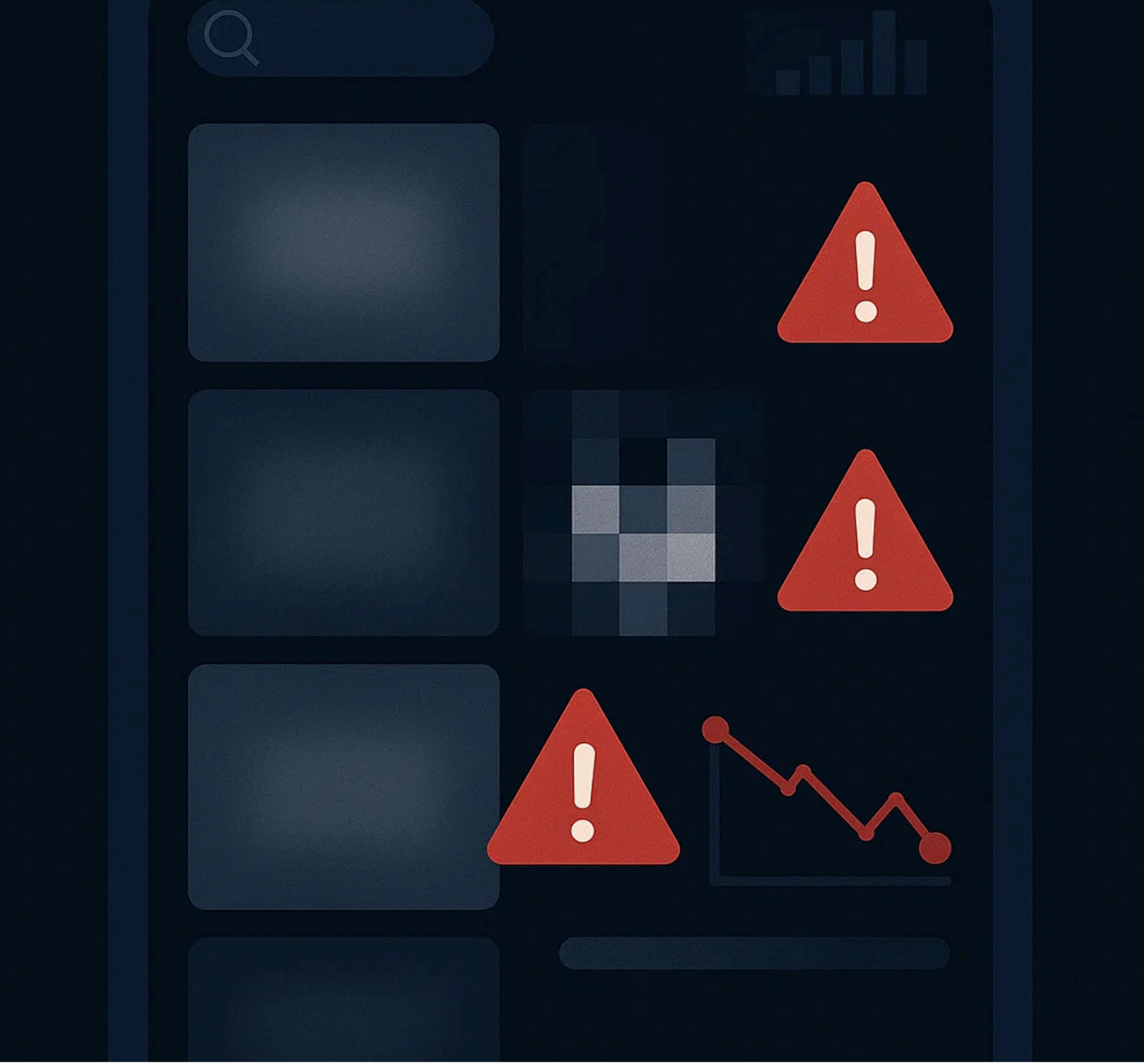

# TIKTOK'S RESEARCH API: PROBLEMS WITHOUT EXPLANATIONS

AI FORENSICS

TikTok API seems to be not fully functional for several months.

# Table of contents

# Credits

**Authors:** Carlos Entrena-Serrano[1], Martin Degeling, Salvatore Romano, Raziye Buse Çetin.

The contribution from AI Forensics is funded by core grants from Open Society Foundations, Luminate, and Limelight Foundation.

**Email:** info@aiforensics.org

---

[1] Centre for Digitalisation, Democracy and Innovation (Brussels School of Governance, Vrije Universiteit Brussel). All other authors are affiliated with AI Forensics.

# Executive Summary

Following the Digital Services Act of 2023, which requires Very Large Online Platforms (VLOPs) and Very Large Online Search Engines (VLOSEs) to facilitate data accessibility for independent research, TikTok augmented its Research API access within Europe in July 2023. This action was intended to ensure compliance with the DSA, bolster transparency, and address systemic risks. Nonetheless, research findings reveal that despite this expansion, notable limitations and inconsistencies persist within the data provided.

- Our experiment reveals that the API fails to provide metadata for one in eight videos provided through data donations, including official TikTok videos, advertisements, and content from specific accounts, without an apparent reason.

- The API data is incomplete, making it unreliable when working with data donations, a prominent methodology for algorithm audits and research on platform accountability.

- To monitor the functionality of the API and eventual fixes implemented by TikTok, we publish [a dashboard](#) with a daily check of the availability of 10 videos that were not retrievable in the last month. The video list includes very well-known accounts, notably that of Taylor Swift.

The current API lacks the necessary capabilities for thorough independent research and scrutiny. It is crucial to support and safeguard researchers who utilize data scraping to independently validate the platform's data quality.

# Introduction

In July 2023, TikTok expanded access to its [Research API to Europe](). While [the company marketed]() this move as a way to facilitate independent research on their platform, the decision was likely motivated by the European Union's Digital Services Act. The DSA and upcoming [data access provision]() under Article 40 require very large online platforms (VLOPs) like TikTok to provide researchers with access to public information.

The research API is only one of several transparency mechanisms required by the DSA. TikTok is also offering a Virtual Computing Environment (VCE) that is made available to non-academic researchers, requiring them to run their queries and data analysis within a coding environment provided by TikTok. According to the documentation, it allows for aggregated analysis on up to 100,000 items and enables analysis on content created by minors, which are excluded from the regular API. The research conducted using the VCE can be monitored entirely by TikTok, and larger queries are manually checked by TikTok employees before the results are available to the researchers.

Another transparency measure is the [Commercial Content Library](), which TikTok made public in July 2023. It lists all advertisements published on the platform targeting users in EU member states. [Researchers have criticized]() the library for being hard to use and for missing important information on commercial posts. Recently, the European Commission issued a [press release]() stating that the advertising library is currently in breach of the DSA, as it fails to disclose the necessary information and offers insufficient search options.

The objective of this project is to test the TikTok official API by querying a set of videos we encountered with different methodologies. The report does not draw any legal conclusions from this analysis, and is not intended to make any accusations against the platform, but rather test the API in a real case scenario.

# Known Limitations of the Research API

We draw from [Daikeler et al. (2024)](#) the following traditionally intrinsic criteria used to assess data quality—criteria focused on the technical properties inherent to the data itself:

- **Accuracy**. The extent to which the data represents reality and is free from errors.
- **Completeness**. The degree to which all necessary data is present and accounted for, with no missing values in the data.
- **Timeliness**. How up-to-date the data is and how quickly it is made available to users.
- **Consistency**. The degree to which data is free from contradictions or conflicts within the same data set or across multiple data sets.
- **Validity**. The extent to which the data is relevant and applicable to the intended use or purpose.

Researchers have raised several concerns regarding the API's **accuracy and completeness**. A [systematic audit](#) found out that (1) the video metadata retrieved from TikTok servers was inaccurate, and (2) some video metadata could not be retrieved from the API, even when it was still publicly accessible on the platform. More than just a simple mistake on TikTok's side, accuracy and completeness issues likely impacted the validity of numerous research projects conducted until TikTok appeared to have "fixed" the problem in September 2024, for example, as in the case of [researchers investigating the 2024 European](#) parliamentary elections, who identified such issues with the TikTok Research API.

There are additional known limitations of the TikTok Research API regarding **completeness** and **timeliness** (which can also impact the validity of research findings). According to TikTok's own documentation, "New videos take up to 48 hours to be added to the search engine"[2] and the video API endpoint will only disclose information on videos that are:

1. made public by a creator who is aged 18 and over;
2. AND, are posted in the regions of US, Europe and Rest of the World;
3. AND do not belong to Canada.[3]

The reason behind TikTok's 48-hour delay for content retrieval via official APIs is unclear, especially since comparable platforms do not impose such restrictions on researcher data access. This limitation could significantly hinder reactive investigations into emerging issues, which require timely access to data to address problems before they worsen. It also makes it impossible to perform some content moderation analysis (which usually occurs within 48 hours) and analysis on virality creation, as the first hours are critical to shaping the reach of content based on its initial performance.

In this report, we conduct further experiments to assess **the completeness of the available data** beyond the known restrictions and test **the consistency of the (non)availability of data**.

# Experiment 1: Data Donation

Our starting point for investigating the current functionality of TikTok APIs comes from our experience as researchers working with them for various projects. More concretely, from using it in combination with TikTok data donations to study users' consumption of content on the platform. This methodological approach begins with a list of TikTok URLs from the donated data and makes API requests to populate metadata for these TikTok videos (Figure 1).

---

[2] See https://developers.tiktok.com/doc/research-api-faq (accessed 20.05.2025)
[3] These limitations are listed in the https://developers.tiktok.com/doc/research-api-codebook in the codebook. Confusingly the "Getting Started" page lists Canadian Videos as an example: https://developers.tiktok.com/doc/research-api-get-started (both accessed 20.05.2025)

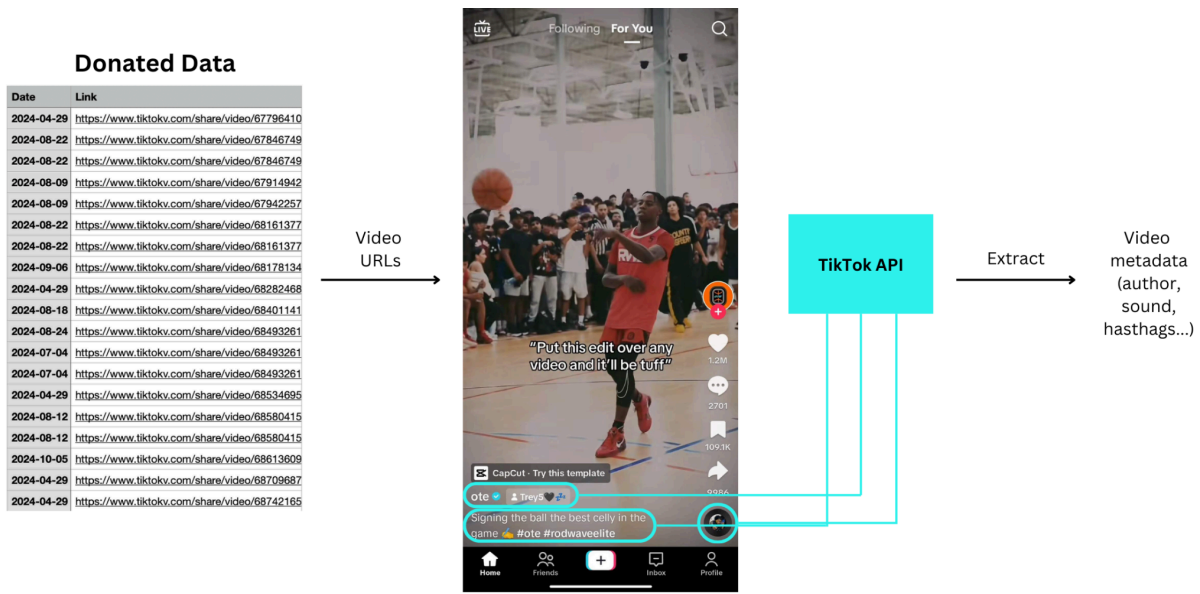

**Figure 1**: *Process of retrieving posts' metadata combining donated data and TikTok's API*

Starting with a concrete list of TikTok posts made it straightforward to check if the API successfully retrieved information from all of them. From an initial sample of approximately 260,000 TikToks, we were initially unable to retrieve metadata for 70,239. TikTok's API did not have informative error codes that specified why some post metadata could not be downloaded (e.g., explaining they had been removed or they were posted by private accounts). At this point, to investigate further, we had to rely on scraping TikTok to check if the unavailable posts were publicly available on the platform.

The research community has [defended](#) scraping as an essential tool to guarantee access to platform data. As the other TikTok API audits and our investigation have shown, it is a fundamental tool that allows researchers and public authorities to confirm the correct functioning and validity of social media platforms' API data. Without scraping, researchers using the API in a different way than us, for example, asking for videos with specific hashtags, have no real way to know if the data returned is accurate or not.

After scraping TikTok, we confirmed that, out of the 70,239 posts, approximately 36% were not public – either deleted, private, or only visible to friends. The other 62.7% were all publicly available on the platform, but not retrievable through the API.

# Reverse-engineering the API

To confirm whether the problem lies in our API query approach, we tested it using different methods. In our initial setup, we requested post IDs in batches of 100, which is the largest size allowed by the API. Using large batches allowed us to more efficiently utilize the 1,000 daily API call limit that TikTok has established. However, at some point during our investigation, we realised that some of these batches had become "corrupted". This meant that if a batch contained IDs for posts that were no longer publicly available, some still-public posts within the same batch were also not returned. To further test this issue, we created new batches of 100 videos and ultimately retrieved around 12,000 posts. Approximately 32,000 TikTok videos remained for testing.

From this point, we began using our daily API calls to request each ID individually, running the same query for two consecutive days. It was a very slow process. Due to the 1,000 daily API call limit, we needed around 64 days to go through the complete list. Following this procedure, we managed to retrieve metadata from some additional TikToks, bringing the total number to 12,373. **Still, we confirmed again that 18961 IDs are still not retrievable from the API**. We interrupted the test on May 16, 2025, to begin writing this report, but we are continuing to recheck the posts left.

# Analysing the non-retrievable content

We were able to identify several distinct features for posts that were not available through the research API. We contacted the TikTok API team and provided examples of content that could not be retrieved. Although we shared our report in advance and agreed to schedule a meeting to discuss the findings, a public explanation of the inconsistencies is still missing. In addition to the publicly disclosed restriction on Canadian content, which we learned about at this point in our research, we found that videos created by TikTok and its regional subsidiaries, as well as those published as ads, were consistently unavailable. Furthermore, we identified a systemic issue wherein approximately 1% of creators had no accessible videos. Below, we will detail each case we found in detail.

## 1) TikTok authorized videos

From time to time TikTok posts company-sponsored videos related to official announcements. This includes videos shown to new users about the risks of challenges, but more prominently, the video that TikTok CEO Shou Zi Chew posted in response to the introduction of the TikTok ban bill in the US Congress with more than 30 million views.

Although these videos are not very common (we identified 32 videos in our dataset of 70k), they are public, oftentimes receive a large number of views, and of course, they are of public interest.

It's unclear why the research API prevents researchers from collecting this type of video. However, we believe it is crucial for platforms to prioritize making their own (political) content accessible to researchers. This is essential for accountability regarding their posted content and sets a positive example.

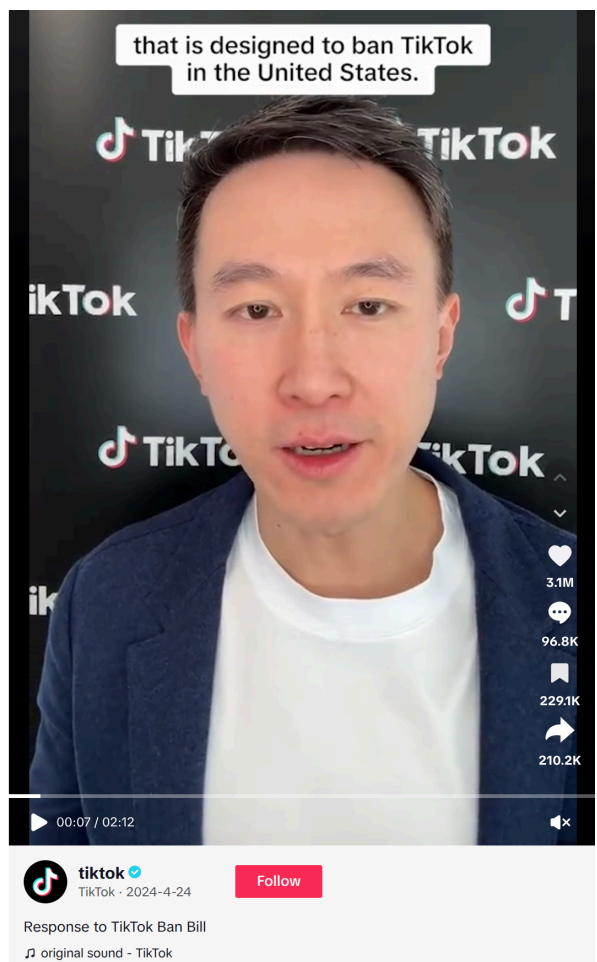

**Figure 2:** *Screenshot of the <u>video</u> by TikTok CEO Shou Zi Chew from April 24th, 2024.*

## 2) Accounts excluded from the API

There are also a number of accounts that seem to be excluded from the research API all together. None of their videos could be retrieved although the majority is publicly available. From our dataset, we could identify 163 of such accounts of which at least 5 different videos were listed, but for which none returned any metadata. Prominent examples include the accounts <u>@askthereddit</u> (3.2 M followers) or the account of the podcaster <u>@lexfridman</u> (600K followers).

From the documentation, we know that the research API does not offer metadata on videos from creators registered in Canada. 90 of the 163 accounts we identified were from creators that listed their region as Canada in the metadata, which leaves 73 accounts where this explanation does not apply. For several of them, further manual research revealed that the creators were in fact Canadian citizens, despite their region being listed differently. It therefore remains unclear to us how TikTok determines an account's location and therefore decides to withhold the data from the research API.

## 3) Advertisements

When ads are shown in the For You Feed, TikTok allows advertisers to hide their real identity behind a username of their choice (Figure 3) or a randomly generated username (Figure 4), accompanied by the label "Sponsored Content".

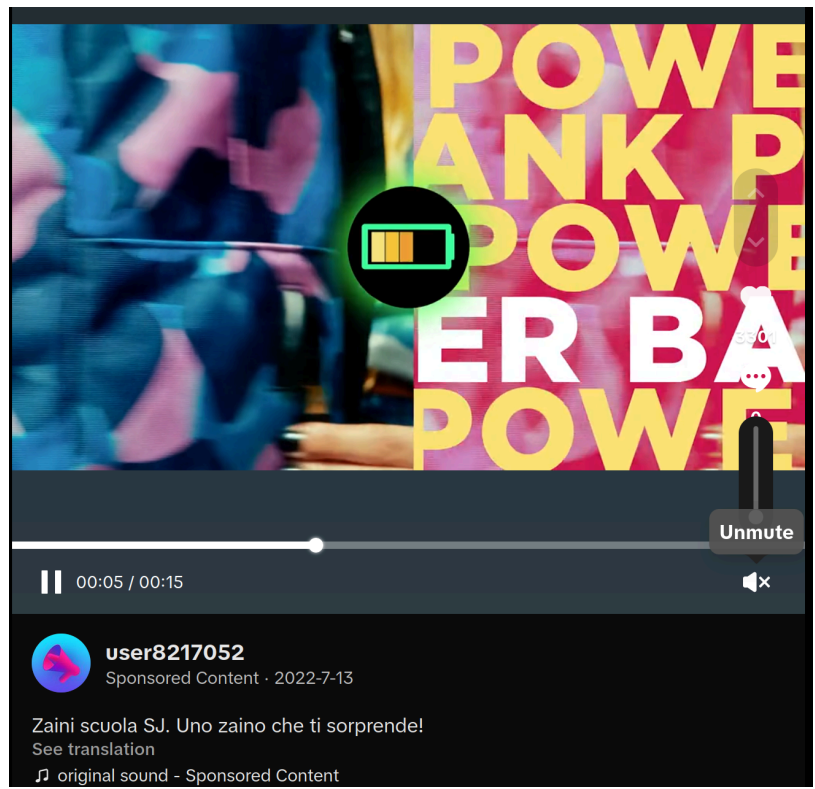

**Figure 3:** Screenshot of a faked ad username on an advertisement that was published on July 14, 2022. Screenshot taken May 8, 2025.

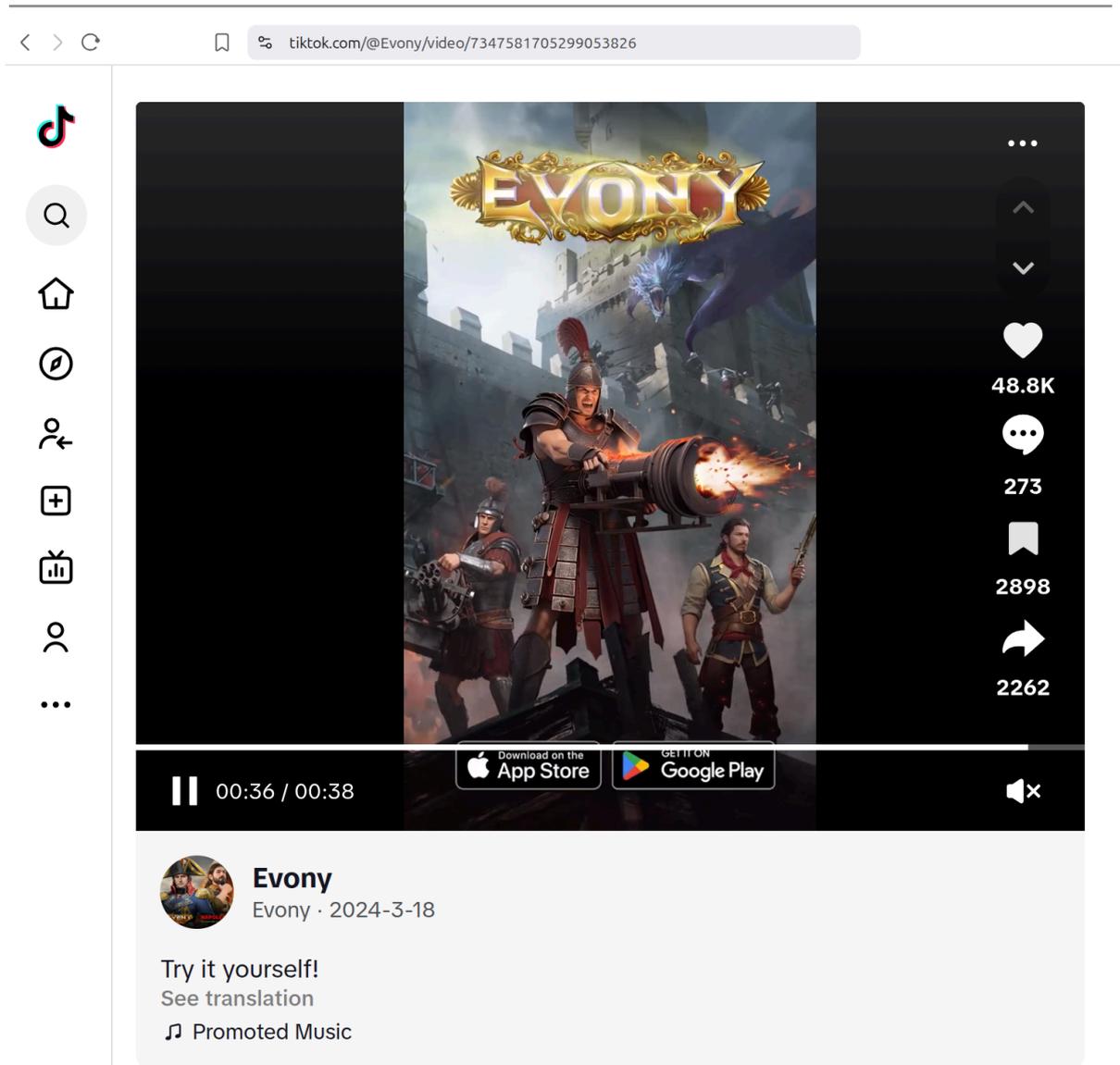

**Figure 4**: Example of an ad for a mobile application that uses the Apps name "Evony" as username, although that username is already taken.

Although the video (Figure 4) exists on the platform, it has a unique identifier and is publicly accessible, the research API does not return any content for that ID when queried.

Additionally, advertisements can use protected sounds, presumably for copyright reasons or to prevent ill-intended remixes of marketing material. In cases where the Music is labeled as "Promoted Music" or is otherwise restricted (Figure 5), the videos, although not involving falsified authors for advertising purposes as described above, are still not retrievable through the research API.

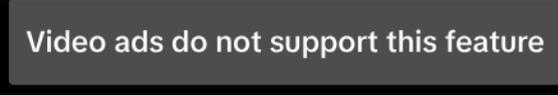

**Figure 5:** *Screenshot of the error message that users get when clicking on a Music that is limited to a specific ad. Screenshot taken from the same video as in Figure 4.*

## Summary of findings

The chart below (*Figure 6*) summarizes our findings by showing what we found about the 70,239 videos we investigated. We were able to successfully retrieve, after multiple attempts, only 18% of the videos for which we missed metadata. Among the 83 percent that remained unavailable, the largest share (36%) of videos can be attributed to the videos being deleted or set to private. This means that the remaining 46% of the videos were public, but not available via the API. While one part of this can be attributed to known limitations of the API (Videos from Canada) we found that a similar share of videos is not available because they were marked as advertisements. For the remaining 21% of videos, we do not know why the API did not return any information, while some might contain minors (another known limitation of the API), a random check did not reveal this as the only reason.

For the original research based on data donations (which collected a total of ≈260k TikToks), this means that roughly 1 in 8 posts (12,46%) of videos could not be analyzed.

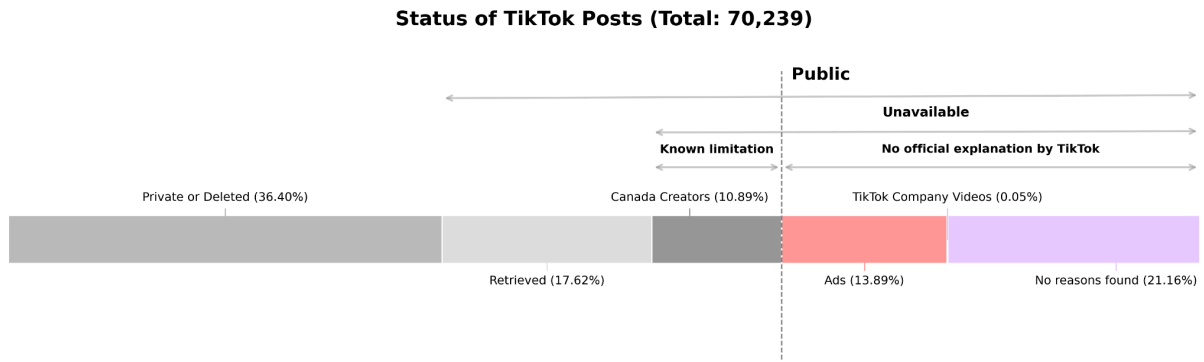

*Figure 6:* Overview of categories of missing/retrievable data from the API based on the initial dataset of non-retrievable posts.

# Experiment 2: Scraping

To assess the extent of this problem more generally with another use case, we collected the metadata of the first 100 videos shown on the German FYP daily for one week (May 14-20, 2025) without being logged in. We then collected the metadata of all videos posted by the creators that were featured, and randomly selected 10 posts of each creator who had posted at least 10 videos, which were at least 3 days old. We used this sample to test their availability on the Research API, given the known limitations. Although the documentation notes that it takes "up to 48 hours" for videos to be available through the research API, we found that the failure rate was high for up to 72 hours, so with a conservative approach, we decided to put our cut off at 72 hours.

As shown in Figure 7, for 69.3% of the creators, we were able to retrieve data from all 10 videos they had posted, while for 1.5%, no video was available. We can confirm that for users where the origin region is set to Canada, no videos are available and have excluded those from the listing below.

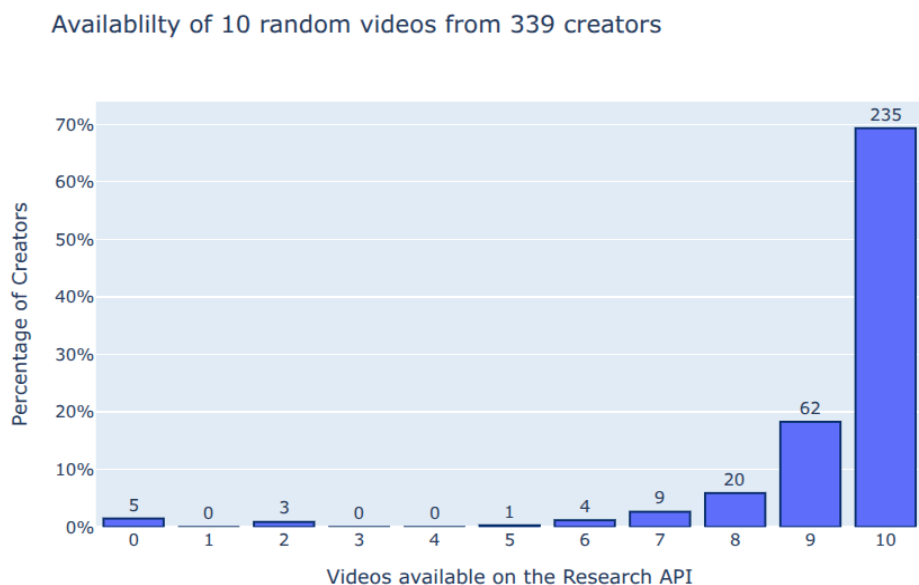

*Figure 7*: *Videos available on the Research API within the first 100 videos on the For You Feed collected daily on May 14-20, 2025. Each bar represents the amount of creators that had a certain number of available videos we could retrieve with the API. The first bar shows that 1.5% of the creators tested had 0 videos available on the API.*

The five creators we encountered for which no video could be retrieved include the news outlet [@thesun](@thesun) (30M followers) or [@barstoolsports](@barstoolsports) (47M followers) which we think should be accessible as they are prominent public profiles.

# Monitoring APIs: a Public Dashboard

TikTok frequently modifies its infrastructure, including its API. These changes have historically resulted in API downtimes, data reporting inaccuracies, or glitches; however, we are not aware of any record of the changes and the issues they addressed on the API. This often leaves the research community in the position of needing to test the tool before it can be used. We decided to create a public dashboard to ensure that the problems we identified are consistent over time. We set up a script to regularly check a set of videos that should be available through the Research API but were not. We went through the lists of videos that were not available, ordered by the number of views. We selected videos with millions of views that matched the categories described above, as well as other videos that were not available, e.g., by creators like Brook Monk (35M followers) or Taylor Swift (32.5M followers). We monitored the availability of 10 selected videos over one month and found that the majority of the videos were consistently not available.

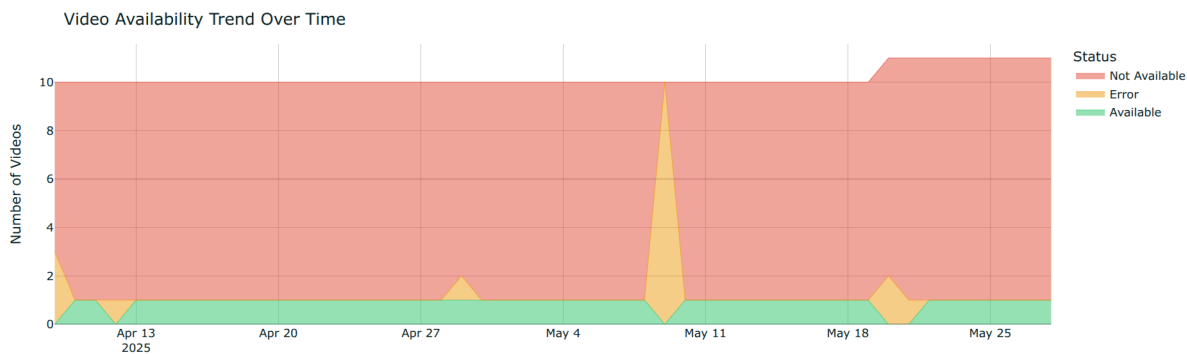

**Figure 8:** *Availability of 10 tested videos in the Public Dashboard. You can check the updated data here.*

We developed the dashboard to enable transparency about TikTok's Research API. We intend to keep the dashboard online to also help researchers understand whether problems they are encountering are affecting only their own account. Our records indicate that certain functionalities have been unavailable since a test conducted in December 2024.

As mentioned above, we will have a call with the TikTok API team to discuss our findings and possibly help the team to fix the issues we and other researchers encountered. We hope to be able to say that all the documentation is updated and complete and that all the data that should be accessible on the official API will actually be available.

In conclusion, our ongoing monitoring efforts are crucial, and we remain committed to working with the relevant teams to rectify the identified issues. We plan to release timely updates regarding any improvements or fixes that have been implemented.

A dashboard of the videos queried daily is available at:

https://playground.tiktok-audit.com/api-na/

# Conclusion

**In sum, our experiments show that TikTok's research API can not serve as a source for research that requires up-to-date, complete, and consistent information.** Especially studies that utilize data donations have to take severe limitations into account, as researchers cannot assess information about all the videos of their participants. Even taking into account the known limitations of recent videos and those published from creators in Canada, our research shows that (1) videos of at least 1% of creators, (2) some TikTok company videos, and (3) almost 10,000 advertisements are not accessible through the API. If TikTok has good reasons to exclude specific videos from the API escapes our assessment, since no justification is transparently disclosed in error messages or otherwise.

Given that TikTok's advertisement library has its limitations, the inability to retrieve ads metadata through the Research API further contributes to the lack of advertising transparency.

**The issues we show are not the first problems with the Research API. Even if TikTok addresses these specific issues, which we urge them to do, a fundamental question remains: can researchers trust the data platforms share through their APIs?** While API access may still be a valuable tool with a lower entry barrier for some researchers, TikTok must acknowledge that it is insufficient and allow other types of data access. For example, researchers should be allowed to use data scraping to access relevant metadata. Data scraping is a technique widely used in the industry, but discouraged in public interest research as it could sometimes be interpreted as a violation of platforms' terms of service. We believe it should be recognised for what it is: a vital tool for transparency.

The European Commission must ensure that TikTok complies with Article 40 of the DSA, providing quality data access through its research API. Furthermore, it is crucial for the European Commission to empower researchers deploying alternative data-gathering tools, such as scraping, and even consider using them in-house for regulatory oversight. Without the ability to independently verify the accuracy of API tools, the promise of data access risks remaining an illusion.